\documentclass[fleqn,twoside]{article}
\usepackage{espcrc2}

\usepackage{graphicx}
\usepackage{amssymb}

\usepackage[figuresright]{rotating}

\newcommand{\AmS}{{\protect\the\textfont2
  A\kern-.1667em\lower.5ex\hbox{M}\kern-.125emS}}

\title{New Way to Compute Excited States and Thermodynamics: 
       Monte Carlo Hamiltonian}

\author{H. Kr\"oger\address[MCSD]{D\'epartement de Physique, 
        Universit\'e Laval, Qu\'ebec, Qu\'ebec G1K 7P4, Canada}%
        \thanks{E-mail address: hkroger@phy.ulaval.ca. 
                Supported by NSERC Canada.},
        X.Q. Luo\address{Department of Physics, Zhongshan University, 
        Guanzhou 510275, China}%
        \thanks{Supported by NSF for distinguished young scientists of China.}
        and 
        K.J.M. Moriarty\address{Department of Mathematics, Statistics 
        and Computer Science, \\
        Dalhousie University, Halifax N.S. B3H 3J5, Canada}%
        \thanks{Supported by NSERC Canada.}
        }
       
\begin{document}

\begin{abstract}
We present a new way to compute thermodynamical observables on the lattice. We compute excited states and thermodynamical functions in the scalar model
via the Monte Carlo Hamiltonian technique.
We find agreement with standard Lagrangian lattice calculations, but observe lesser fluctuations in the results from the MC Hamiltonian.  
\vspace{1pc}
\end{abstract}

\maketitle

\section{MONTE CARLO HAMILTONIAN}

The idea of the Monte Carlo Hamiltonian has been introduced in Refs.\cite{Jirari99,KleinGordon01,Salzburg01,Huang02}. Here we want to recall the basic idea, outline some of its characteristic features and present some selected results. Comparing the Lagrangian approach with the Hamiltonian approach in lattice field theory, the Hamiltonian approach has not 
flourished in producing a wealth of numerical data like the Lagrangian approach. On the other hand, there are a few topics, where the Lagrangian approach has only given slow progress, e.g. on excited states, wave functions, particle scattering. In our opinion, the lattice Hamiltonian approach might have done a lot better, if it were not for the lack of effective many-body methods to solve such Hamiltonian and compute physical observables. Recall that the Kogut-Susskind Hamiltonian of lattice $QED_{3+1}$ and $QCD_{3+1}$ is a quantum mechanical operator in a high-dimensional Hilbert space and a numerical many-body solution is quite difficult.

\begin{figure}[thb]
\vspace{9pt}
\begin{center}
\includegraphics[scale=0.3,angle=270]{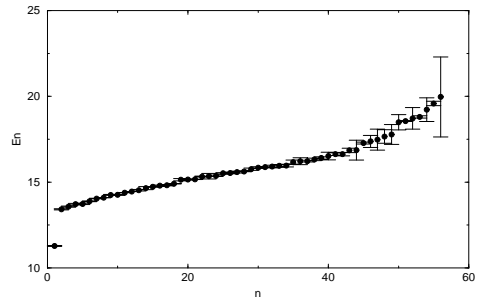}
\end{center}
\caption{Energy spectrum in a low energy window.}
\label{Fig1}
\end{figure}

The Monte Carlo Hamiltonian using a stochastic basis is an attempt to address this problem. Let us outline the idea. First, the Monte Carlo Hamiltonian is not an operator of canonical form. It is rather defined in a subspace of Hilbert space and is given in terms of matrix elements. The physical idea is that this object plays the role of an effective Hamiltonian. It has a window of validity.
This can be visualized as a window in the energy spectrum. Beyond this window, the effective Hamiltonian is unphysical.
When computing thermodynamical functions, the window of the energy spectrum translates into a temperature window. Such a window can be observed in the numerical data (see below). The notion of a window is familar in physics from the concept a scaling window and also reminds of Wilson-Kadanoff's renormalisation group, where a Hamiltonian is constructed which gives the correct physics of critical phenomena  (at phase transitions), but is unphysical otherwise.

\begin{figure}[thb]
\begin{center}
\includegraphics[scale=0.3,angle=270]{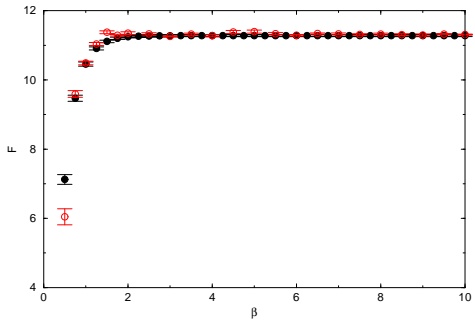}
\end{center}
\caption{Free energy $F({\beta})$. 
Monte Carlo Hamiltonian (full dots) vs. standard Lagrangian lattice calculations
(open dots).}
\label{Fig2}
\end{figure}

What is the guiding principle in constructing such effective Hamiltonian?
In contrast to the transfer matrix formalism which defines a Hamilton operator 
in the limit $a_{t} \to 0$, we start by considering physical amplitudes corresponding to finite transition time $T$. The finiteness of $T$ is in part responsable for the finite energy window. However, in contrast to Lagrangian lattice field theory, where one considers n-point functions corresponding to vacuum-to-vacuum transitions, we go beyond the vacuum state and consider transition amplitudes from other states which probe in part the wave functions of excited states. To give an example in quantum mechanics, we consider transition amplitudes $G(x_{f},t=T;x_{i},t=0)$, where $|x_{i}>$ and $|x_{f}>$ denote eigen states of initial and final position. We compute those quantities via standard path integrals on the lattice using Monte Carlo with importance sampling (the trick is to write the path integral as a ratio). In order to get a reasonable representative subspace of Hilbert space (which also influences the size of the window), we took in quantum mechanics $x_{i}$ and $x_{f}$ from a finite grid of position states and computed all possible combinations of transitions. Then an algebraic diagonalisation of such a matrix yields a finite set of eigenstates and eigen energies of the effective Hamiltonian, i.e. 
the Monte Carlo Hamiltonian. For test cases from quantum mechanics this has been shown to work \cite{Jirari99}.

\begin{figure}[thb]
\begin{center}
\includegraphics[scale=0.3,angle=270]{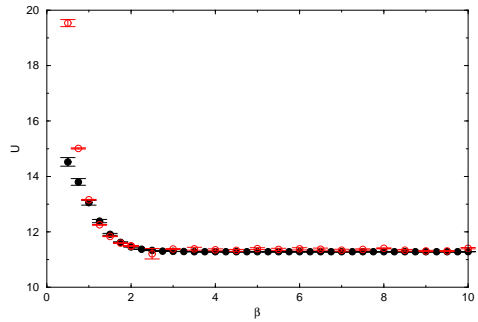}
\end{center}
\caption{Same as Fig.\ref{Fig2}, for average energy $U({\beta})$.}
\label{Fig3}
\end{figure}

\section{MANY-BODY SYSTEMS AND STOCHASTIC BASIS}
It is simple to convince oneself that a straight forward generalisation to
many-body systems and field theory is not feasible. The solution is to 
select physically important degrees of freedom. Here we use as guiding principle the enormous success of Lagrangian lattice field theory, and in particular the successful working of Monte Carlo with importance sampling. In Lagrangian lattice field theory it is possible to estimate an observable by taking the mean of the observable over a relatively small number of "equilibrium' field configurations, which have been constructed using a weightfactor built from the action.

\begin{figure}[thb]
\begin{center}
\includegraphics[scale=0.3,angle=270]{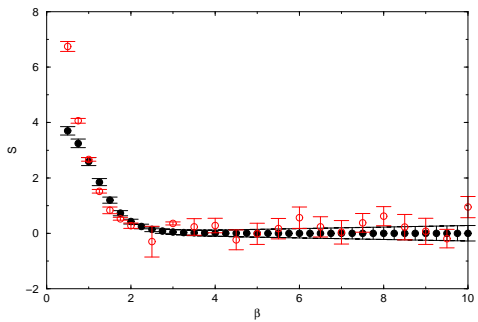}
\end{center}
\caption{Same as Fig.\ref{Fig2}, for entropy $S({\beta})$.}
\label{Fig4}
\end{figure}

We suggest that 'equilibrium' field configurations in Lagrangian lattice field theory have a close analogy in Hamiltonian lattice field theory in terms of states of a stochastic basis. In other words, taking 'equilibrium' configurations at a fixed time slice yields Hilbert states (Bargman states) which represent physically important degrees of freedom. This constitutes the so-called stochastic basis. Thus, via Monte Carlo with importance sampling, involving the weightfactor with the action, we generate a subspace of Hilbert space of a relatively small dimension (in the order of 100 to 1000), which 
represents the important degrees of freedom for the physics in a certain window.  

The demonstration that such repesentative basis exists and allows to compute 
excited states and thermodynamical functions has been first made for the chain of coupled harmonic oscillators (Klein-Gordon model). In Ref.\cite{KleinGordon01,Salzburg01} the excitation spectrum has been compared with the analytic answer, establishing the existence of an energy window and thermodynamic functions (specific heat) have been computed.

\begin{figure}[thb]
\begin{center}
\includegraphics[scale=0.3,angle=270]{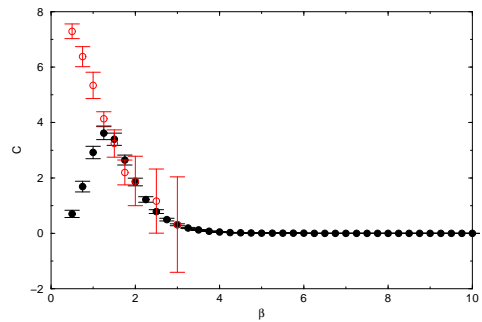}
\end{center}
\caption{Same as Fig.\ref{Fig2},  for specific heat $C({\beta})$.}
\label{Fig5}
\end{figure}

The method has been extended to the scalar model in Ref.\cite{Huang02}. 
This model is not analytically soluble, hence we have chosen to compare our results with standard Langian lattice calculations. Fig.\ref{Fig1} presents the the first 50 eigenvalues of the excitation spectrum of the Monte Carlo Hamiltonian. A sudden increase of errors indicates the upper limit of the energy window. The thermodynamical functions free energy, average energy, entropy and specific heat are displayed in Figs.\ref{Fig2}-\ref{Fig5}. The temperature window reaches from temperature zero to an upper temperature given by $\beta \approx 1 - 1.5$. The size of the window depends on transition time $T$, the size of the stochastic basis, statistical errors, etc. 
In Figs.\ref{Fig2} - \ref{Fig5} one observes that the data from the Monte Carlo Hamiltonian show lesser fluctuations than those from the standard Lagrangian approach.

\end{document}